\documentclass[a4paper]{spie}  

\usepackage[]{graphicx}
\usepackage{amssymb,amsmath,amsfonts}
\usepackage{wrapfig}
\usepackage{txfonts}

\title{Astronomical photonics in the context of infrared interferometry and high-resolution spectroscopy}

\author{Lucas~Labadie\supit{a}, 
Jean-Philippe~Berger\supit{b},
Nick~Cvetojevic\supit{c},
Roger Haynes\supit{d},
Robert Harris\supit{e},
Nemanja Jovanovic\supit{f},
Sylvestre~Lacour\supit{g},
Guillermo~Martin\supit{h},
Stefano~Minardi\supit{d,i},
Guy~Perrin\supit{g},
Martin~Roth\supit{d}
Robert~R.~Thomson\supit{j}
\skiplinehalf
\supit{a} I.\,Physikalisches Institut, Universit\"at zu K\"oln, Z\"ulpicher Str. 77, 50937 K\"oln, Germany \\
\supit{b} ESO, Karl-Schwarzschild-Str. 2, 85748 Garching bei M\"unchen, Germany\\
\supit{c} MQ Photonics Research Centre, Department of Physics and Astronomy, Macquarie University, NSW 2109, Australia\\
\supit{d} Leibniz-Institut f\"ur Astrophysik Potsdam, An der Sternwarte 16, D-14482 Potsdam, Germany\\
\supit{e} Zentrum f\"ur Astronomie der Universit\"at Heidelberg, Landessternwarte K\"onigstuhl, K\"onigstuhl 12, 69117 Heidelberg, Germany\\
\supit{f} Subaru Telescope, National Astronomical Observatory of Japan, Hilo, HI 96720, USA\\
\supit{g} LESIA, Observatoire de Paris, PSL Research University, CNRS, Sorbonne Universit\'es, UPMC Univ. Paris 06, Univ. Paris Diderot, Sorbonne Paris Cit\'e, France\\
\supit{h} Univ. Grenoble Alpes/CNRS, IPAG, F-38000 Grenoble, France\\
\supit{i} Institute of Appl. Phys., Friedrich-Schiller-Universit\"at, Max-Wien-Platz 1, 07743 Jena, Germany\\
\supit{j} Scottish Universities Physics Alliance (SUPA), Institute of Photonics and Quantum Sciences, Heriot-Watt University, Edinburgh, EH14 4AS, United Kingdom}

\authorinfo{Further author information: (Send correspondence to L.Labadie)\\L.Labadie: E-mail: labadie@ph1.uni-koeln.de, Telephone:\,+49 221 470 3493
}

\begin{document} 
  \maketitle 

\begin{abstract}
We review the potential of Astrophotonics, a relatively young field at the interface between photonics and astronomical instrumentation, for spectro-interferometry. We review some fundamental aspects of photonic science that drove the emergence of astrophotonics, and highlight the achievements in observational astrophysics. We analyze the prospects for further technological development also considering the potential synergies with other fields of physics (e.g. non-linear optics in condensed matter physics). We also stress the central role of fiber optics in routing and transporting light, delivering complex filters, or interfacing instruments and telescopes, more specifically in the context of a growing usage of adaptive optics. \end{abstract}



\section{Introduction}


Imaging and spectroscopy form the foundation of 
observational astrophysics. Since early times, astronomical instrumentation has made use of optics principles to collect, focus and disperse stellar light and implemented increasingly complex opto-mechanical designs that are now found on the most advanced large facilities. 
Important discoveries in astronomy always came after a major step in the development of new instrumentation and thanks to synergies with other fields of physics. This has been the case, for instance, with new concepts of optical slicers enabling integral-field spectroscopy\cite{Iserlohe2004}\,, with highly impacting advances in the field of detector physics to develop more sensitive and faster focal plane arrays\cite{Jorden2014,Guieu2014}\,, or more recently with the rapid growth of adaptive optics system to compensate for the degrading effect of atmospheric turbulence\cite{Davies2012}\,. \\
Photonics has also emerged as a possible breakthrough driver in astronomical instrumentation\cite{Bland-Hawthorn2009}\,. While it is probably too early to appreciate its long-term impact in ground- and space-based astronomy, the new field of astronomical photonics has grown because of a demand in astronomy to solve new problems that had arisen with the advent of larger facilities, and in which the classical approach of bulk optics instruments has not been fully satisfactory. \\
What is {\it astrophotonics}, or photonics in the context of astronomical instrumentation? Broadly speaking it is the discipline that deals 
with the transport, processing and shaping of stellar (or non-stellar) radiation by mean 
of {\it waveguides}, although the term may cover also other applications such as micro- and nano-optics, phase control devices, or light generators for calibration purposes\cite{Bland-Hawthorn2009}. 
Even before the expression was coined, 
the use of optical fibers in multi-object spectroscopy led since the early eighties to a growing interest in photonics, which was reflected in the {\it Fiber Optics in Astronomy} conference series published in 1988\cite{Fiber1}\,, 1993\cite{Fiber2}\,, 1998\cite{Fiber3} and revisited in 2014. Since then, more elaborated photonics functions have been imported in astronomy and (astro)photonics instruments designate now those designs in which the core optical functionality is ensured by small-scale photonics devices. The integration of miniaturized optical functions in real instruments is now a fact and this approach is being increasingly recognized and followed. Conversely, bulk optics instruments refer to more conventional designs involving mirrors, lenses, windows, and dispersing elements such as prisms and gratings. Their size and cost typically increase accordingly with the telescope size. The boundaries astrophotonics are not firmly set and it is likely that in the coming years the field will evolve with new applications in fields such as astrometry, metrology or high time resolution astronomy. The potential impact of astrophotonics in the current panorama of instrumentation is twofold:
\begin{itemize}
\item fostering new approaches for the design of large instruments less sensitive to environmental disturbances (temperature, vibrations...) based on improved packaging and miniaturisation, as well as facilitate incremental updates of instruments by reducing their costs and complexity of usage.
\item enabling new functionalities only attainable with photonics solutions to uniquely expand performance of existing facilities
\end{itemize}
While the term astrophotonics covers in principle the whole electromagnetic spectrum, it is in our context more specifically associated with astronomical applications in the visible and infrared range, which build-up on the telecommunications and solid-state physics fields. Nonetheless, a comparable problem of beam shaping, mixing, and instrument miniaturisation is being considered for instrumentation in the far-infrared and sub-millimeter range. This will however not be addressed here.
\\
\\
In this paper we attempt a comprehensive review on the advances in optical and infrared astrophotonics, which concern primarily the fields of interferometric imaging and high-resolution spectroscopy. These two fields have advanced so far independently on the astrophotonics scene but with potential synergies in the future. Astrophotonics has already demonstrated strong potential at the VLTI and with the arrival of new large facilities like the ELTs astrophotonics is able to feed new ideas in the way of designing instruments.

\section{Requirements overview}

\subsection{Science requirements}
The basis of any astronomical instrument being with the scientific observations, a number of top-down science and instrument requirements that apply also in the case of a photonic-based approach need to be considered. Some of these requirements are reminded hereafter.
\begin{itemize}
\item Astronomy is a field where observing programs are inherently photon-starved. Therefore high throughput is the first requirement to be fulfilled for any photonics-based instrumental concept. It encompasses the requirements on the intrinsic transparency of the material, on the propagation losses of the guiding structure (which depend on the achievable waveguide opto-geometrical parameters) and on the input and output coupling efficiency within a turbulent environment. 
Quantitatively, this requirement is very much depending on the targeted application: at the Cassegrain or Nasmyth focus of a telescope the stellar flux has been attenuated by few tens of percent at most. In the case of long-baseline interferometry, the flux collected down at the interferometric lab after traveling along the delay lines is about few percent due to the numerous interfaces through which light is transmitted or reflected. 
\item The operational band(s) and bandwidth is also a top requirement, directly connected to the science driver of the instrument. Ideally, an instrument would cover several astronomical bands (e.g. with the instruments NACO and MATISSE at the VLT/VLTI). For programs like deep imaging it is also important that the instrument performs well over the entire wavelength band of interest.
\item Science cases which goal is to accurately trace gas lines and kinematics or precisely follow the Doppler shift of stellar lines due to an orbiting planetary companion require high spectral resolution, typically in the range of $R\sim$5\,000 to 100\,000 (the case of ultra-high spectral resolution via heterodyne spectroscopy is not addressed here) and a very stable instrumental transfer function. For instance, reaching an accuracy of $\sim$1\,m/s on a modern spectrograph with a resolving power of $R$=200,000 requires a precision of 1/1,000 element. This implies that any external spectral noise contributions have to be extremely well identified and suppressed.
\item Achieving the diffraction-limited resolution of the telescope and high Strehl ratios is very important when the science case focuses on the study of small-size and highly contrasted objects like, for instance, very low-mass companions or disk structures close to a much brighter primary component. Whereas sub-milliarcsecond resolution is sought, one typically requires adaptive optics or even single-mode long baseline interferometry with baselines of the order of hundred meters.
\item The science cases that focus on understanding galaxy evolution and cosmology usually need to access a wider field of view with multiplexing capabilities, i.e. the ability to pick-up simultaneously hundreds of different angular directions in the field of view.
\end{itemize}

\subsection{Short reminder on important properties of waveguides}

\begin{figure}[b]
\centering
\includegraphics[width=0.75\textwidth]{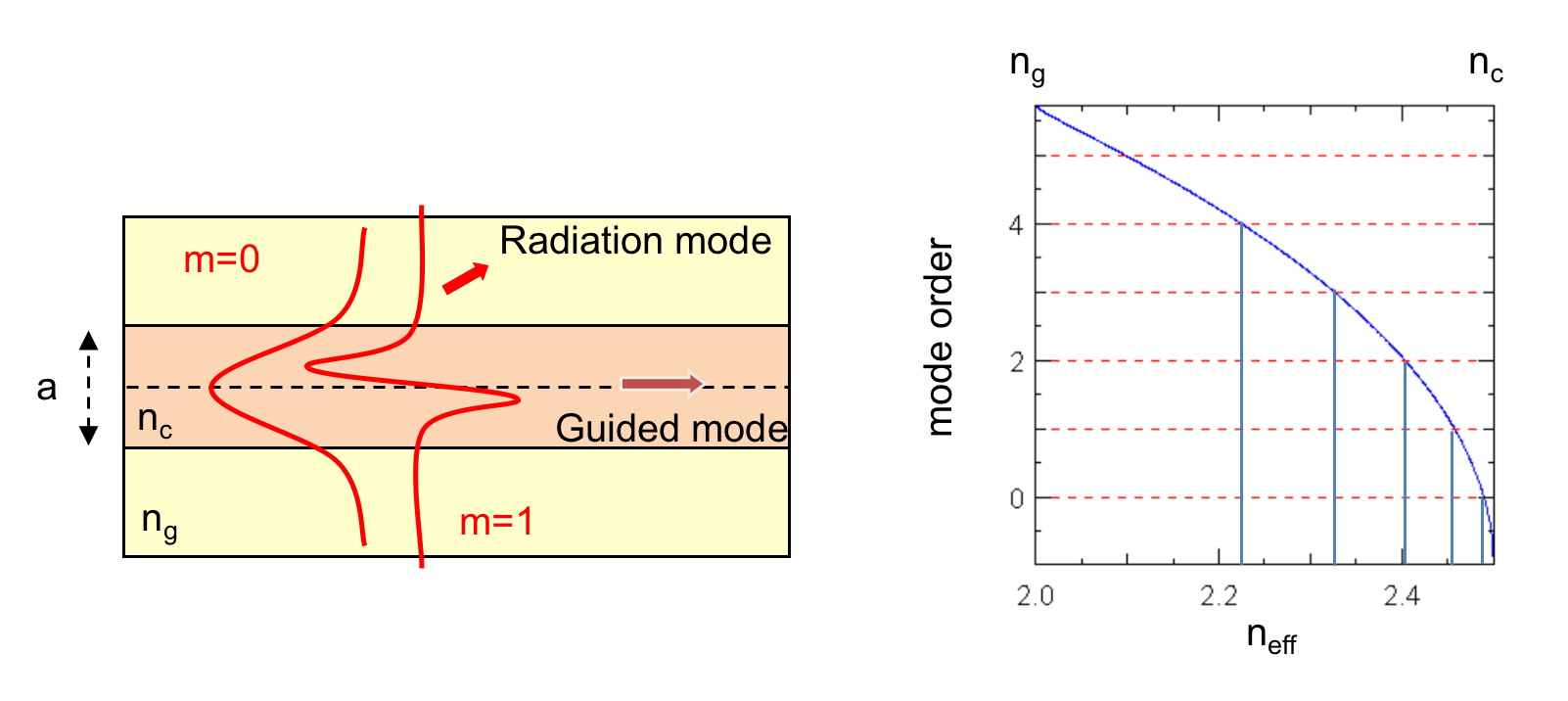}
\caption{Left: sketch of a symmetric waveguide geometry with the transverse amplitude distribution of mode 0 and 1. Right: example of numerical solving of the dispersion equation for a waveguide with diameter $a$=4\,$\mu$m, $n_{\rm c}$=2.5, $n_{\rm g}$=2.0 and $\lambda$=2\,$\mu$m.}\label{fig1}
\vspace{0.0cm}
\end{figure}

We briefly remind some fundamentals of guided optics. A waveguide is typically formed of a high index core surrounded by a low-index cladding (or substrate). The condition of total internal reflection at the core/cladding interface geometrically determines a maximum acceptance {\it half}-angle $\theta_{\rm 0}$\footnote{The relation is valid for a multimode waveguide} given by $n_{\rm 0}.$sin($\theta_0$)=$(n_{\rm c}^2-n_{\rm g}^2)^{1/2}$, where $n_{\rm c}$, $n_{\rm g}$ and $n_{\rm 0}$ are, respectively, the refractive index of the core, of the cladding and of the surrounding medium (typically air). From geometrical considerations on the constructive state of the propagating wave, a dispersion equation can be derived, which gives the effective index $n_{\rm eff}$ associated to each propagation mode. In the case of a simple symmetric slab waveguide, this dispersion relation is given by
\begin{eqnarray}
k_{\rm 0}.d.\sqrt{n_{\rm c}^2-n_{\rm eff}^2}-2.\tan^{-1}(g.\frac{\sqrt{n_{\rm eff}^2-n_{\rm g}^2}}{\sqrt{n_{\rm c}^2-n_{\rm eff}^2}})&=&m.\pi \label{eq1}
\end{eqnarray}
\noindent The wave vector is $k_{\rm 0}$=2$\pi$/$\lambda$ and $d$ the core diameter. The parameter $g$ depends on the considered polarization TE or TM\cite{Labadie2006OpEx}\,. The effective index $n_{\rm eff}$ is interpreted as the index {\it seen} by the corresponding propagation mode. It only depends on the opto-geometrical parameters of the waveguides $d$, $n_{\rm c}$, $n_{\rm eff}$, and on the operating wavelength $\lambda$. Solving Eq.~\ref{eq1} provides the solutions for the spectrum of effective indices, with the increasing ones corresponding to the modes of decreasing order. The effective index is linked to the {\it mode propagation constant} by $\beta_{\rm n}$=(2$\pi$/$\lambda$)$n_{\rm eff,n}$, which is the fundamental parameter to quantify the behaviour of waveguides that is involved in the design of a photonic-based instrument.

\subsection{Instrument requirements for a photonic-based instrumental design}\label{Sect22}

The above mentioned science requirements translate into instrument requirements to ideally define and design the most suited instrument. In the real world of astrophotonics this top-down approach is only be partially followed and the achievable science is also tailored to some extent to the actual capabilities of existing photonic components. This has nonetheless led to important science breakthrough.
\begin{figure}[t]
\centering
\includegraphics[width=0.5\textwidth]{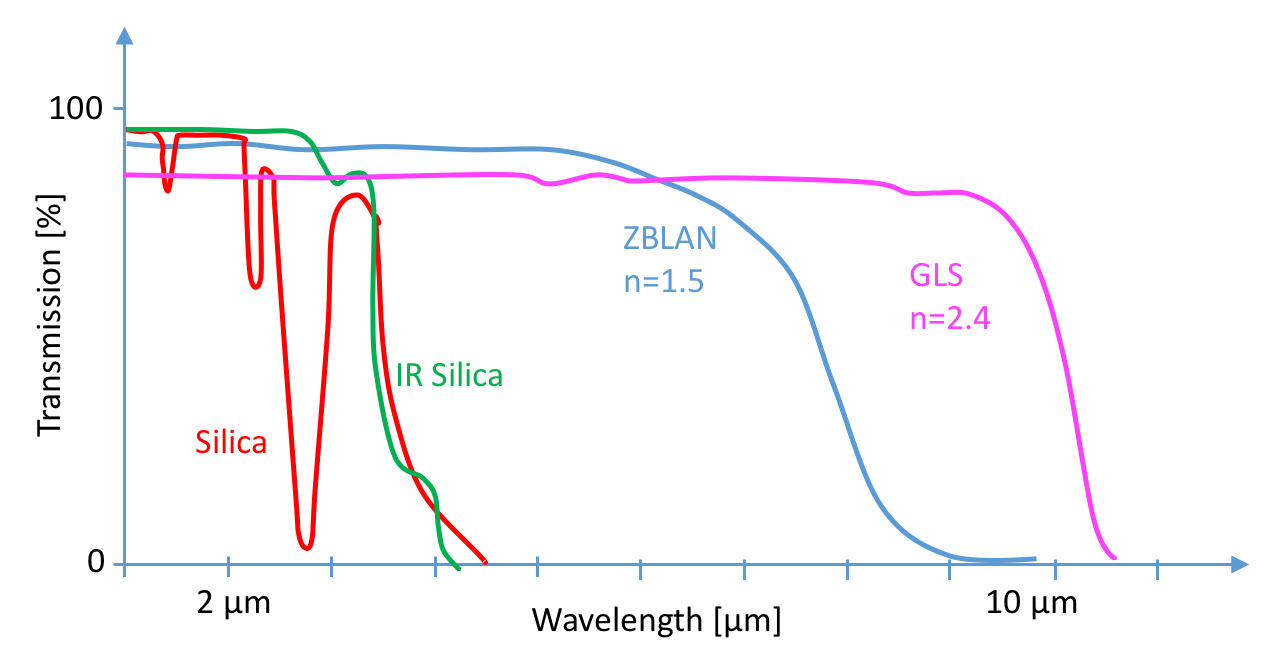}
\caption{\label{fig1}{Transparency of some IR materials in the near- and mid-infrared range.}}
\vspace{0.0cm}
\end{figure}
\begin{itemize}

\item {\bf Transparency range:} With respect to science requirements 1 and 2, the material transparency range is a key technical requirement to consider at first. At visible and near-infrared wavelengths, fused silica is the most commonly used substrate thanks to its excellent transparency. The properties of this chemically stable and non-toxic glass are very well established. Its low refractive index ($n\sim$1.4) ensures low Fresnel reflections of the bulk material ($\lesssim$\,5\% for front and back facets).  The OH content of fused silica leads to strong absorption bands beyond 2.5\,$\mu$m. This can be overcome with low-OH ``dry" IR grade silica like {\it Infrasil}, which then presents higher transmission up to $\sim$3.5$\,\mu$m. for short lengths.\\
Operating at longer wavelengths can in principle be achieved by selecting any of the numerous infrared glasses that cover the infrared range up to 20--30\,$\mu$m. Other aspects need to be considered such as chemical and mechanical stability, potential level of toxicity, temperature working range compatible with cryogenic operations, intrinsic material birefringence, no excessive fragility and low aging. The majority of infrared glasses present a high refractive index (up to $n\sim$4.0 for Germanium), which induces high Fresnel reflection losses. 
However the selection of an apparently suitable substrate does not guarantee the feasibility of waveguides or photonic functions suitable for astronomy as this is fully dependent on the technological platform. Currently, assessed or promising materials compatible with an existing waveguide technological platform are fused silica for the visible and near-infrared spectral range (up to 2.4\,$\mu$m), Lithium Niobate and fluoride glasses like e.g. ZBLAN with high transparency up to $\sim$5 and 6.5\,$\mu$m respectively, and chalcogenide glasses like e.g. GLS (Gallium Lanthanum Sulphide) transparent up to $\sim$9\,$\mu$m. 
\begin{figure}[t]
\centering
\includegraphics[width=0.75\textwidth]{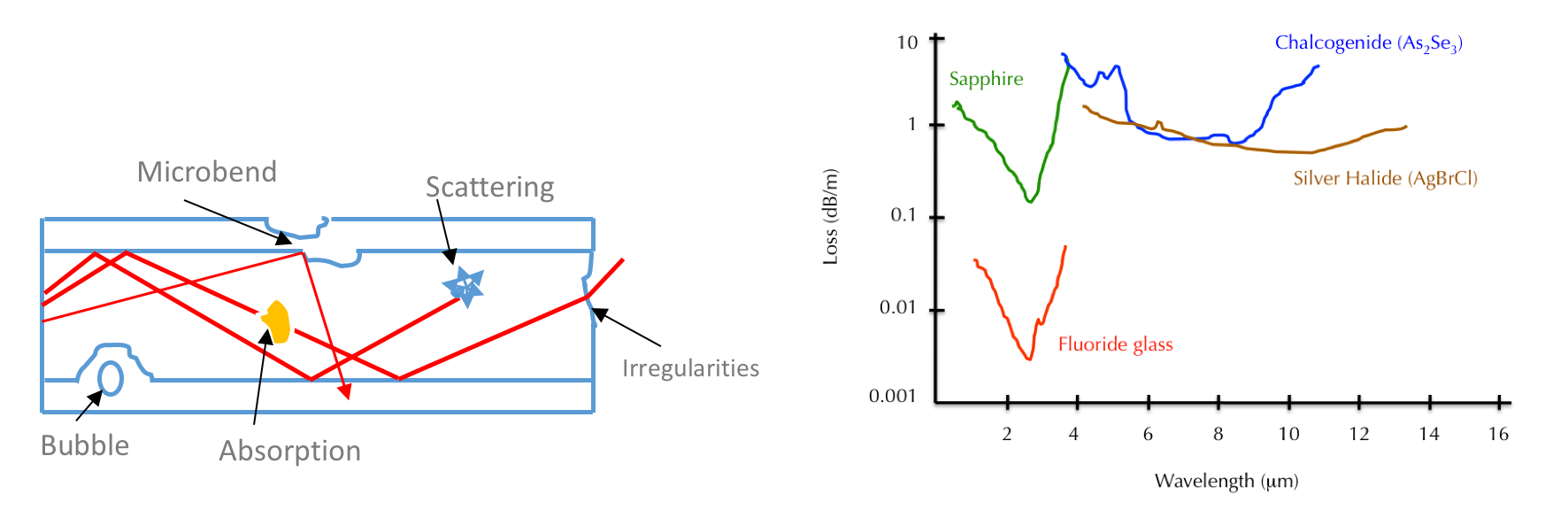}
\caption{\label{fig2}{Left: schematic view of the different sources of extrinsic losses in a waveguide. Right: transparency range of some mid-IR fibers. Adapted from Harrington (2010).}}\label{Fig2}
\vspace{0.0cm}
\end{figure}
\item {\bf Propagation losses:} While intrinsic losses of the chosen material can be very low (e.g. $<$0.01 dB/km for Silica, $<$0.005 dB/km for ZBLAN), in practice extrinsic losses due to impurities and scattering centers arising during the manufacturing process will dominate in a waveguide or a fiber (cf. Fig.~\ref{Fig2}). The total propagation losses can easily reach few dB/km for fluoride fibers, which suggests that the transparency spectrum of a waveguide might be very different from bulk it is made of. It is interesting to notice how the manufacturing of glasses and waveguides in micro-gravity can significantly improve the purity of the sample (e.g., see Tucker (1997)\cite{Tucker1997}\, for the case of heavy metal fluoride fibers).
\noindent Minimizing the propagation losses does not only depend on the manufacturing process. It is also strongly depends on the design of the guiding structure in which bending losses as well as radiation into the waveguide's cladding may play an important role (see below).

\item {\bf Single-mode or multimode operation:} Operating a single-mode instrument may be implemented in applications where excellent wavefront quality is required, which translates into an accurate measurement of the V$^2$ visibilities in long-baseline interferometry or in the removal of modal noise for high-resolution spectroscopy. We remind that in a single-mode waveguide the transverse amplitude and phase of the propagated wave is uniquely determined by the opto-geometrical parameters of the guiding structure, independently of the input coupling conditions. Such a waveguide acts as a modal filter suppressing any phase information contained in the coupled field. Vice-versa, highly multimode waveguides will be used to couple and collect most of the telescope light arising in the form of an uncorrected or partially corrected point-spread-function (PSF). Single-mode versus multimode operation is critically linked to the science goal involved and in particular to the high-flux or low-flux regime involved. Tatulli et al. (2010)\cite{Tatulli2010}\, makes interestingly the case for long-baseline interferometry. The cutoff frequency of a waveguide is given by
\begin{eqnarray}
V&=&\frac{2\pi}{\lambda}a\sqrt{n_{c}^2-n_{g}^2} \label{eq2}
\end{eqnarray}
which separates the single-mode (V$<$2.4) from the multimode regime (V$>$2.4). The term $a$ is here the waveguide's core radius. From the above equation, it is understandable that single-mode waveguides must have small core radius assuming standard refractive index contrasts. In the case of the well-known telecom fibers at 1.3\,$\mu$m, single-mode cores are about 5--10\,$\mu$m in diameter with a 125\,$\mu$m cladding diameter, whereas the core of the multimode fiber is almost ten times larger, with a standard radius of 62.5\,$\mu$m. 

\item {\bf Mode field diameter and field confinement:}
\begin{figure}[b]
\centering
\includegraphics[width=0.75\textwidth]{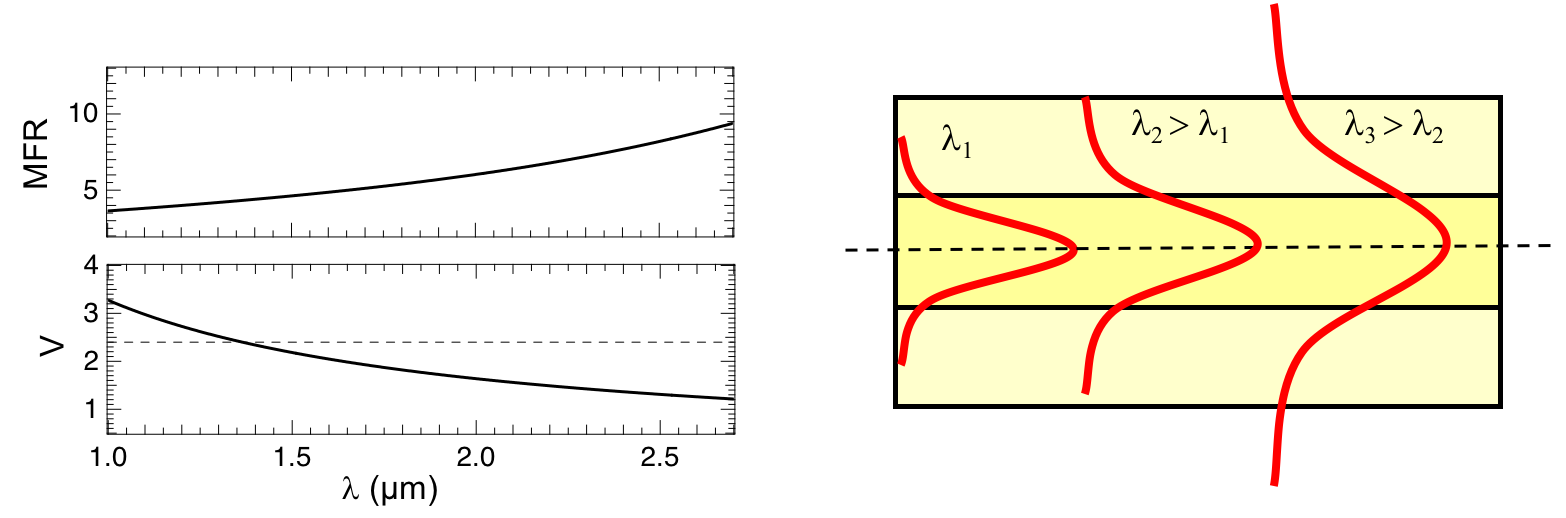}
\caption{\label{fig2}{Left: trends for the $V$ parameter and mode field radius as a function of wavelength. The horizontal dashed line shows the limit between multimode and single-mode regime. Here the index contrast is assumed to remain constant. Right: Visualization of the lower confinement with increasing wavelength.}}\label{Fig3}
\end{figure}
A good approximation of the {\it fundamental} mode field radius $w$ is given by the Marcuse's equation 
\begin{eqnarray}
w&=&a\times(0.65+\frac{1.619}{V^{3/2}}+\frac{2.879}{V^{6}})\label{eq3}
\end{eqnarray}
where $V$ is the aforementioned normalized frequency, $a$ again the core radius. Eq.~\ref{eq2} suggests that for a waveguide with a given core radius $a$, $V$ will generally decrease with increasing wavelength -- assuming that index contrast $\sqrt{n_{c}^2-n_{g}^2}$ remains roughly constant with $\lambda$ --  or with decreasing index contrast at constant wavelength. In both cases the mode field diameter 2$w$ will increase according to Eq.~\ref{eq3} and result in poorer confinement of the field (see illustration of Fig.~\ref{Fig3}). The consequences are a larger fraction of the field in the structure being radiated away in the cladding, a decrease of the effective index of a given mode, and therefore a rising of the bending losses. The last effect can be particularly limiting for designing compact integrated optics functions since tighter bends will unavoidably lead to larger net losses for a given quality of the waveguides.\\
A stronger confinement of the field in the waveguide -- obtained with a large refractive index contrast -- will diminish the bending losses on the one hand, but on the other hand will complicate the coupling into the waveguide. 

\item {\bf Coupling into a waveguide:} All photonics-based devices deal with the problem of coupling stellar light into a waveguide. \\
In the case of long-baseline interferometry where single-mode fibers are used, poor coupling can quickly become a killing factor if not assisted by some-level of tip-tilt compensation or adaptive optics correction\cite{Shaklan1988}\,. 
For a fundamental mode field radius $w$, a good approximation of the half-divergence angle is given by $\theta\sim$tan$^{-1}$$(\lambda/\pi w)$, which can be connected to the familiar f-number in optics through $f/D$\,$\sim$\,$\pi w/2\lambda$. For instance, the core diameter of a single mode fiber is around 5\,$\mu$m and the fundamental mode field diameter (MFD) is about 40\% larger for a typical $V$=1.8, which would require a beam with numerical aperture f/\#$\sim$7 or faster. 
Also, the size of the telescope PSF has to be roughly of the same order and its relative motion controlled to a fraction of it. Unless operating with telescopes of the size of the Fried parameter r$_{\rm 0}$ or less at the wavelength of observation, coupling in single-mode devices should ideally be assisted by adaptive optics to improve the instrument total throughput. In this sense, this is why the 1-m and 1.8-m telescopes of CHARA and the VLTI respectively will be equipped with adaptive optics in the coming years. Note that alternative geometries to the classical step-index waveguide such as Large Mode Area (LMA) fibers and Photonics Crystal Fibers (PCF) could be used to enlarge the MFD in combination to slower coupling optics, ensuring at the same time single-mode guidance. But none of these solutions have been implemented in astronomy so far.\\
In spectroscopic applications, coupling occurs predominantly between a seeing-limited PSF and a multimode fiber with core diameters as large as 900\,$\mu$m to transport light and feed the spectrometer front-end. In the multimode regime, the resulting MFD is very comparable to the core diameter, hence slow beams typically delivered at the telescope are well suited and the problem of the coupling in this case is relaxed. The need for adaptive optics is also less critical. However, multimode operation may bring other issues relevant for high-precision spectroscopy.

\item {\bf Modal noise and spectral resolution:} Multimode fibers are commonly used in fiber-fed spectroscopy where a partially corrected or seeing-limited point spread function matches well with the fiber large core. However, the illumination pattern of a multimode fiber is by definition not uniform as all supported modes are potentially contributing to this pattern due to the constant and rapid change of the coupling conditions and stresses on the fiber. This results in the so-called {\it modal noise}, which introduces variable spectral line shifts, in other words spectral noise. With the accuracy on the radial velocities expected for detecting Earth-like planets, such a noise must be suppressed as much as possible. Mechanical mode scramblers have been implemented in astronomy so far. Other options rely on adiabatically transforming a multimode input into multiple or extended single-mode outputs are being considered by different groups. These options employ multicore fibers\cite{Haynes2014} and photonic lanterns\cite{LeonSaval2005,Thomson2011}\,.

\item {\bf Waveguide birefringence:} A single-mode waveguide or fiber actually supports two modes that have identical amplitude profiles but orthogonal polarizations referred to as TE and TM modes. The same applies to high-order mode in multimode fibers. In an ideally perfectly circular-core and homogeneous waveguide, these two TE and TM modes would have exactly the same propagation constant (or phase velocity) $\beta$=$(2\pi/\lambda)$$n_{\rm eff}$\,. In the real world, imperfections in the core shape and refractive index homogeneity induces slightly different propagation constants between the two modes, resulting in the scrambling of the input polarization 
and random cross-talk between the two modes. For optical waveguides, birefringence is typically quoted as the difference between the effective indices of the TE and TM modes. Broadly speaking, two aspects are involved in explaining birefringence in waveguides. \\
{\bf \underline {Optical birefringence:}} Most of classical waveguides (i.e. with no polarization-maintaining properties)  does not exhibit birefringence by nature as they are isotropic medium. However, inhomogeneities and anisotropic stresses resulting from the fabrication process introduces unwanted {\it random} residual birefringence also called stress-induced birefringence, which can be interpreted as polarization-dependent refractive index. In the case of optical {\it fibers}, additional time-variable random birefringence can result from external factors such as temperature or mechanical stresses (e.g. bend, twist). This last effect is very much reduced in integrated optics functions.\\
{\bf \underline {Geometrical birefringence:}} Beside the optical birefringence, perfectly isotropic waveguides for which the core cross section lacks azimuthal symmetry would exhibit some level of {\it geometrical} or shape-birefringence (see Fig.~\ref{Fig4}). For instance, if we consider a putative single-mode buried waveguide with a 2$\times$2\,$\mu$m square cross-section with a core index n$_c$=1.55, $\Delta$n=0.1 with the cladding, and calculate at $\lambda$=1.55\,$\mu$m the effective index for the TE and TM fundamental mode, we find a slight birefringence of $\Delta$n$_{\rm eff}$=2$\times$10$^{-5}$ (and not zero since the core is not circular). For the same waveguide but a rectangular buried core with dimensions 5$\times$1\,$\mu$m, the geometrical birefringence raises to $\Delta$n$_{\rm eff}$=4$\times$10$^{-3}$.\\
The total birefringence is the combined result of these two effects. In application like interferometry where polarization mismatches between the different channels need to be minimized, the {\it differential} birefringence is the quantity to be controlled rather than the absolute birefringence itself in each waveguide. In that sense, integrated optics exhibit lower lever of random birefringence with respect to fiber X-couplers. In spectroscopic applications, birefringence in multimode fibers is not a critical issue since usually a maximum scrambling of the state of polarization is sought.
\begin{figure}[t]
\centering
\includegraphics[width=0.95\textwidth]{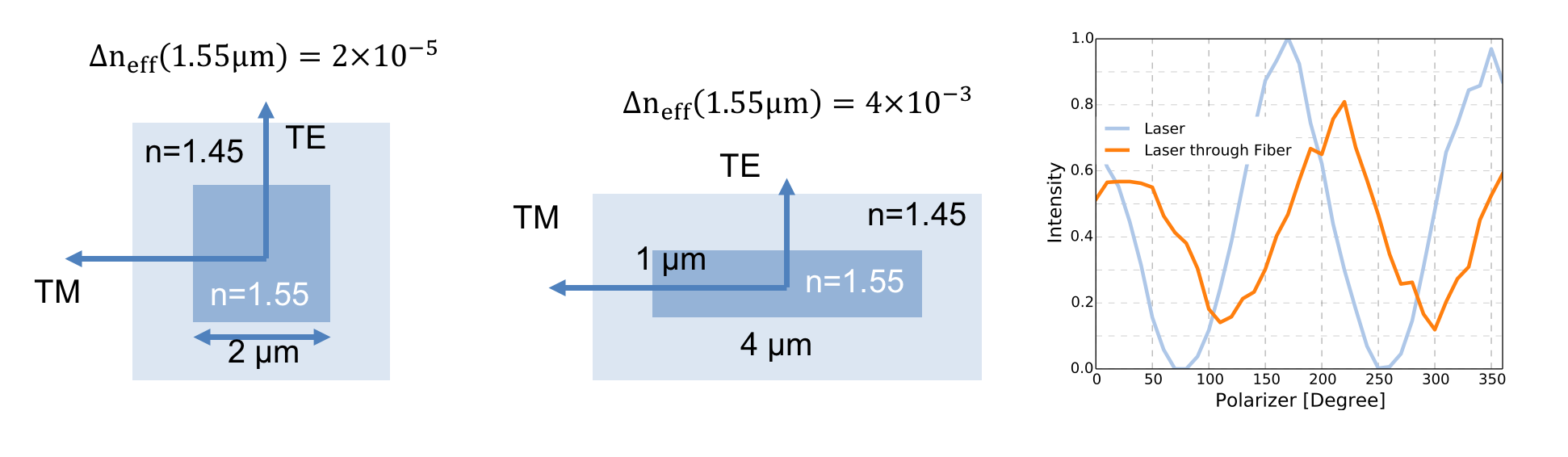}
\caption{\label{fig2}{Left and middle: principle of shape birefringence with its corresponding numerical estimate $\Delta$n$_{\rm eff}$. Right: polarization scrambling in a commercial mid-IR fluoride fiber.}}\label{Fig4}
\end{figure}

\item {\bf Dispersion:} In a waveguide a mode is fully characterized by its effective index or its propagation constant $\beta$=$k_{\rm 0}.n_{\rm eff}$. The wavelength-dependence (or frequency-dependence) of this propagation constant is at the origin of dispersion. The meaningful quantity here is the group velocity dispersion $D_{\lambda}$, or dispersion parameter, given by
\begin{eqnarray}
D_{\rm \lambda}&=&-\frac{2\pi.c}{\lambda^2}\frac{d^2\beta}{d\omega^2}
\end{eqnarray}
In our context, two contributions to the dispersion can be mainly identified. A first one is the material dispersion $D_m$ that is classically explained with the wavelength-dependence of the refractive index, which usually decreases with increasing wavelength. A second one is the so-called waveguide dispersion  $D_w$, which directly results from the fact that we are dealing with propagation modes in a guiding structure. Even in the absence of material dispersion the dependence of the mode confinement on wavelength results in the same dependence of the propagation constant for either multimode or single-mode structures. Hence the statement that modal dispersion only occurs in multimode waveguides is incorrect. 
To first order, the total dispersion $D_\lambda$ is the sum of the contributions $D_m$ and $D_w$ and is given in ps/km.nm. These two contributions may compensate each other at a particular wavelength, leading to zero-dispersion waveguides in a narrow spectral range, a property exploited in long-distance optical communications.\\
Dispersion mitigation is an important requirement in interferometry. However it is the {\it differential} dispersion between the two (or more) interferometric arms that matters rather than the absolute value itself. Differential dispersion results in the phase curvature of the recorded interferogram, which then spreads out significantly and may result in a dramatic loss of interferometric contrast. Following the formalism of Coud\'e du Foresto (1995)\cite{Foresto1995}\,, the phase curvature can be approximated by
\begin{eqnarray}
\frac{d^2\Phi}{d\sigma^2}&=&-\frac{2\pi c}{\sigma^2}(D_{\lambda}\Delta.L+L.\Delta D_{\lambda})
\end{eqnarray}
where $\Delta L$ and $\Delta D_{\lambda}$ are the differential length and the differential dispersion between the arms, respectively, and $L$ and $D_{\lambda}$ are the average dispersion and length, respectively. The differential dispersion can be qualitatively interpreted as how much the manufacturing process may result in slight differences in the optical properties between the two arms, and will typically dominate the phase curvature for centimeter-size integrated optics components or fiber X-couplers. Ideally, of both differential terms are nulled, the phase curvature equals zero. In practice, if the phase curvature can be kept as low as $\sim$2$\times$10$^{-5}$\,rad.cm$^{-2}$, the interferogram can be considered as moderately dispersive.
\end{itemize}

\subsection{Available technological platforms in astrophotonics}\label{platform}

\begin{figure}[t]
\centering
\includegraphics[width=0.95\textwidth]{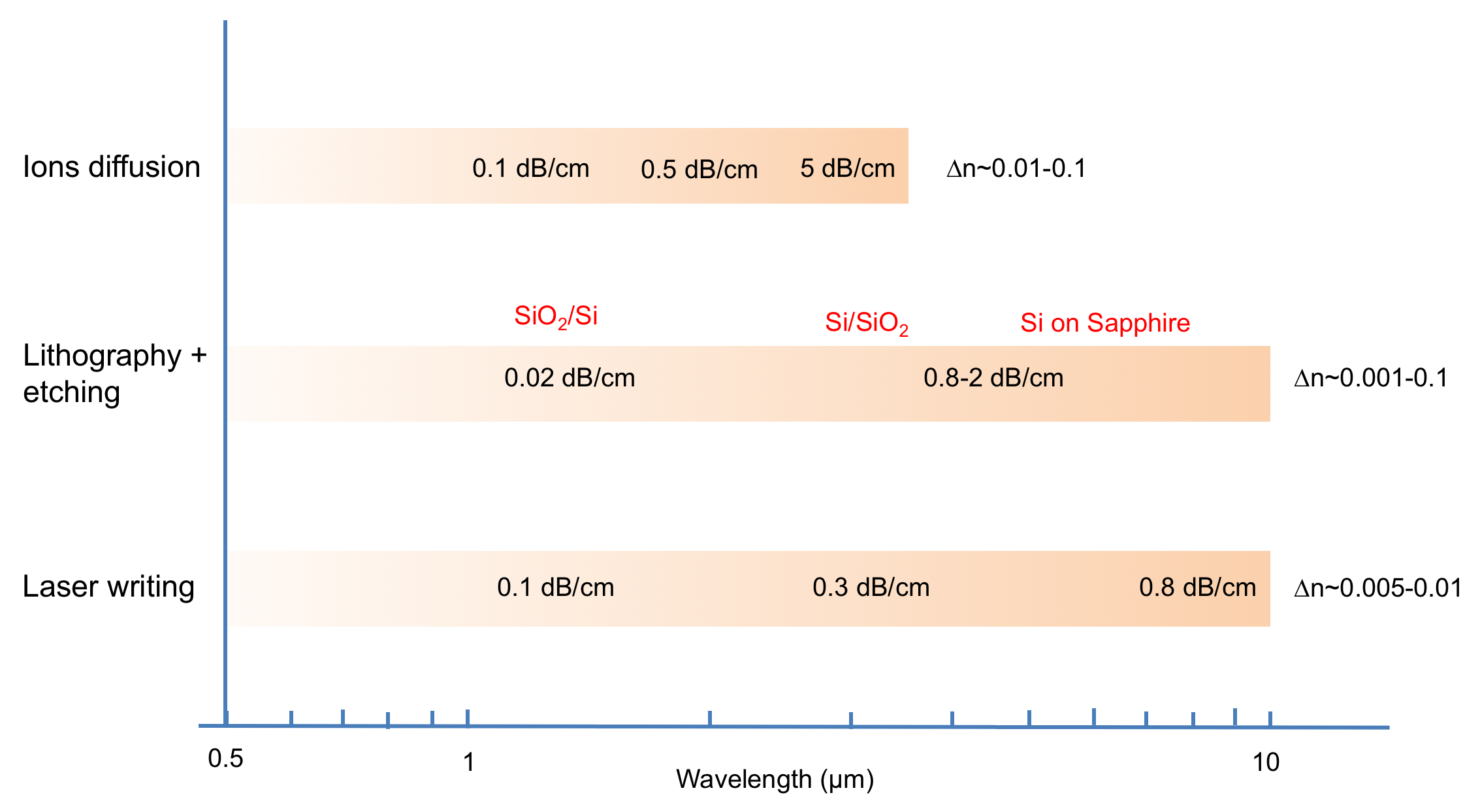}
\caption{Properties and performances of three different technological platforms: Ions diffusion\cite{Broquin2001}; Lithography and etching\cite{Labeye2008}; Laser writing\cite{Ho2006,Gross2015}.}\label{Fig6}
\end{figure}

The important requirements for a platform are the achievable index contrast $\Delta$n that impacts the level of field confinement, the achievable resolution on $\Delta$n, the minimum and maximum achievable waveguide size and its corresponding accuracy, the suitable materials that the platform can process.\\
So far, mainly three technology platforms have been considered to develop astrophotonics components. Fig.~\ref{Fig6} summarizes the performances of the ions diffusion, of the lithography and etching, and of the laser writing techniques. The wavelength coverage, the typical propagation losses and achievable index contrasts are reported. The first two technologies have been used extensively to produce interferometric beam combiners\cite{Laurent2002,Labeye2008}\,, including the one for the GRAVITY instrument at the VLTI. The figure also reports the interesting case of the silicon-on-insulator\cite{Mashanovich2011} and of the silicon-on-sapphire\cite{Li2011} platforms to extend the use of lithography and etching to longer wavelengths.
\\
Recently, the technology of laser writing in glass has been implemented with success in the field of astronomical instrumentation for interferometric and spectroscopic applications. Two approaches coexist. 
One approach is the two-dimensional laser writing, in which a simple CW laser is focused at the surface of a layer for which the bandgap energy of the material corresponds to the photon energy. If the photon is sub-bandgap for the underlying substrate, then photons will only be absorbed by the surface layer under high irradiance nonlinear regime and result in a locally confined increase of the refractive index\cite{Ho2006,Labadie2012}\,. 
A second approach makes use of femtosecond lasers able to concentrate few hundreds of kW to few MW power peaks in glass regions of $\sim$10\,$\mu$m$^3$ at MHz repetition rate\cite{Thomson2009,Gross2015}\,. The high peak powers involved in this technique called ultrafast laser inscription (ULI) enables multi-photon absorption by the glass, which similarly to the CW laser writing approach results in a local increase of the refractive index.\\
Both 2D and 3D writing technique allow the manufacturing of guiding structures by translation of the sample under the laser beam. In addition, because the ULI approach applies to bulk substrates, the focalised laser spot can be translated vertically and inscribe three-dimensional structures that avoid the crossing between waveguides as opposed to the two-dimensional case\cite{Rodenas2012}\,.\\

\begin{figure}[t]
\centering
\includegraphics[width=0.95\textwidth]{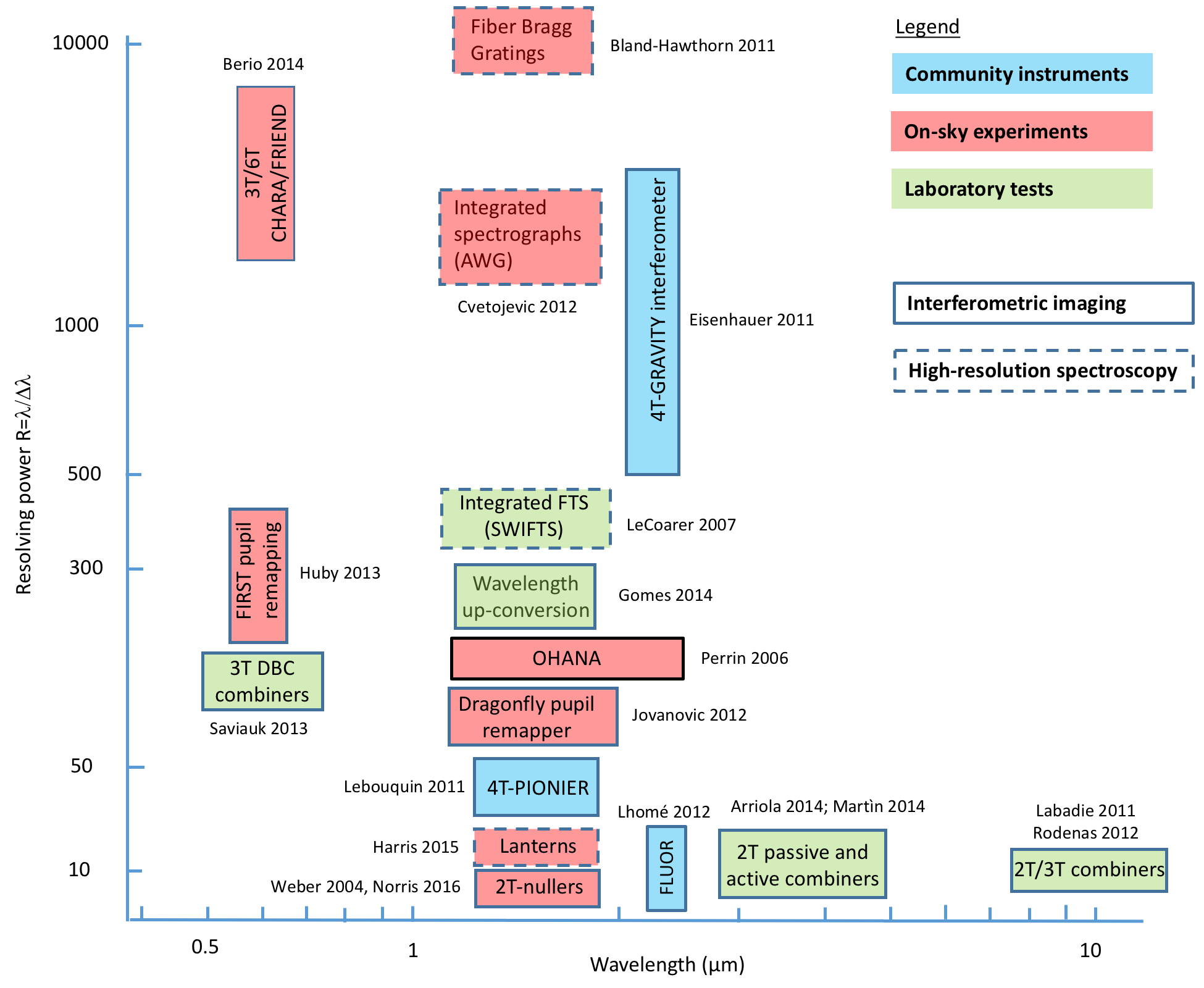}
\caption{General view of the current landscape in astrophotonics instrumentation at optical, near- and mid-infrared wavelengths. See text for explanations. For sake of clarity, multiaxial instruments in which only wavefront filtering by mean of single-mode fibers is implemented are not reported here (e.g. AMBER/VLTI, MIRC/CHARA). References are given at the end of the paper, except for FLUOR\cite{Lhome2012}\,, wavelength up-conversion\cite{Gomes2014} and FRIEND\cite{Berio2014}.}\label{Fig7}
\end{figure}

\section{Current landscape}

Since the first works on fibered interferometers\cite{Froehly1981} and fiber-fed spectrographs\cite{Fiber1}\,, the landscape of photonic-based instruments and prototypes has significantly evolved and in several directions. 
The graph of Fig.~\ref{Fig7} attempts to summarize the distribution of photonic-based initiatives in a wavelength versus spectral resolution plot. In the horizontal axis, instruments and projects are coarsely classified according to the astronomical band of operation, namely the visible (0.6--0.8\,$\mu$m), the near-infrared with the H (1.4--1.8\,$\mu$m) and K (2-2.5\,$\mu$m) bands, the mid-infrared with the L and M bands (3--5\,$\mu$m) and the N band (8--12\,$\mu$m). The vertical axis gives approximatively the spectral resolution at which a given experiment or instrument is operating, either because it is placed in front of a conventional spectrograph or because light dispersion is achieved by the photonics device itself . The plot does not make a difference between operation at high-spectral resolution over a larger bandwidth (e.g. GRAVITY) or operation in a single narrow-band filter. For sake of clarity, it does not display neither high-dynamic range nulling results where singlemode fibers or waveguides have been used only for the purpose of wavefront filtering\cite{Haguenauer2006,Labadie2007,Buisset2007}\,, nor pure technological development of optical fibers.\\
In the current landscape of astrophotonics instrumentation the plot identifies under different colors three categories, namely open community instruments in the scientific exploitation phase, on-sky experiments in the advanced assessment phase, and laboratory experiments focused on the technological development and optimization. Finally, a distinction is made between works focusing primarily on interferometry or spectroscopy, respectively.\\
One obvious remark suggested by this plot is the significant concentration of instrument projects in the astronomical H-band with respect to the other wavelength ranges, which is naturally explained by the large and low-cost availability of silica photonics devices and corresponding technological platforms backed by fifty years of R\&D in the field of optical telecommunication. Such a level of maturity is for instance not reached for photonic devices operating at longer wavelengths. Furthermore, any photonic-based instrument has to deal soon or later with coupling stellar photons into a waveguide. Because the more stringent coupling conditions at shorter wavelengths due to the atmospheric turbulence, coupling is easier to achieve in the H band than in the visible, in particular for single mode instruments. The advent of better performing adaptive optics systems for the visible spectral range should in the future allow new photonics instruments to emerge in this wavelength range.

\section{Examples of photonic devices for astronomy}

As mentioned earlier, most of photonic-based initiatives in astronomy are found as applications for interferometric imaging and high-resolution spectroscopy. 
It is interesting to note that to date mainly on-sky experiments with visitor instruments involving advanced photonics chips have been conducted so far by various international teams. To the best of our knowledge, GRAVITY and PIONIER at the VLTI are the only community instruments based on a photonics-core.  

\subsection{Long-baseline interferometry}

The FLUOR instrument\cite{Foresto1997} can be considered as the first photonics interferometer. It is based on the implementation of fiber X- and Y-couplers to obtain simultaneously the interferometric signal and the photometric channels for calibration.  
An improved version of the fibered approach came with the first implementation of near-infrared integrated optics with the IONIC experiment at the IOTA interferometer (see Kern (1997)\cite{Kern1997}\,, Malbet (1999)\cite{Malbet1999} and references herein). IONIC was first a two-telescope interferometric beam combiner, followed by a three-telescope version delivering first closure phase results on spectrosopic binaries\cite{Monnier2004}\,. The beam combiners were manufactured using standard silica technological platforms described in Sect.~\ref{platform}. The integrated optics (IO) route brought unprecedented compactness, stability and calibration possibilities in the field of interferometry.\\
In preparation to the second generation instruments for the VLTI, a follow-up of these pioneering works was the single-mode four-telescope integrated optics type of beam combiner that later equipped PIONIER\cite{Lebouquin2011} and GRAVITY\cite{Eisenhauer2011} at the VLTI (see Fig.~\ref{Fig8}). The added value of the integrated optics design was to enable the simultaneous measurement of four phase states in quadrature (also known as ABCD scheme\cite{Colavita1999}) for the derivation of visibility amplitudes and closure phases. 
A unique feature of the design enabled by the photonic nature of the device is the phase-shifting function, which ingeniously exploits the control of the effective index of the propagating mode by adequately shaping the cross-section of the waveguide\cite{Benisty2009}\,. Based on the silica-on-silicon technology, the total throughput is as high as $\sim$70\% across the K band, with a slightly decreasing transmission between the cut-off of 2.1\,$\mu$m.\\
The fringe encoding scheme also has a direct impact on the resulting SNR of the detected fringes and different all-in-one and pairwise IO designs have already been analyzed and compared\cite{Lebouquin2004}\,. 
A novel approach is for instance the discrete beam combiners (DBC\cite{Minardi2010}), a combination concept based on the properties of light propagation in 2D periodic arrays of straight, evanescently coupled waveguides. DBC require an array of at least $M=N_t^2$ waveguides to measure all possible mutual coherence functions of the input beams from the measurement of the output excitation pattern. The simultaneous combination of 4-, 6- and 8- telescopes has been demonstrated numerically\cite{Diener2016,Errmann2016}, and first prototypes of 3-telescope DBC have been tested in the laboratory with monochromatic\cite{Minardi2012} and polychromatic\cite{Saviauk2013} light in the R-band. DBC show no waveguide cross-overs or bending and make them potential candidates for high-throughput IO devices. Other concepts rely on a multi-axial multimode interferences (MMI)\cite{Rooms2003} design. 
The possibility to extend the benefit of integrated optics to longer wavelength beyond the silica cutoff has been explored intensively\cite{Labadie2012}\,. The need for a mature and well-understood technological platform capable of dealing with infrared glasses is the key. In the last years, several elementary two-telescope and three-telescope building blocks have been successfully developed for mid-infrared operation\cite{Labadie2011,Rodenas2012,Arriola2014,Tepper2016}\,, in particular thanks to the advent of ultrafast laser inscription in astronomy (see Fig.~\ref{Fig8}). For the near-future, the manufacturing of a mid-infrared integrated optics beam combiner optimized for on-sky operation is the next goal to be achieved.

\subsection{Pupil remapping}

\begin{figure}[t]
\centering
\includegraphics[width=0.48\textwidth]{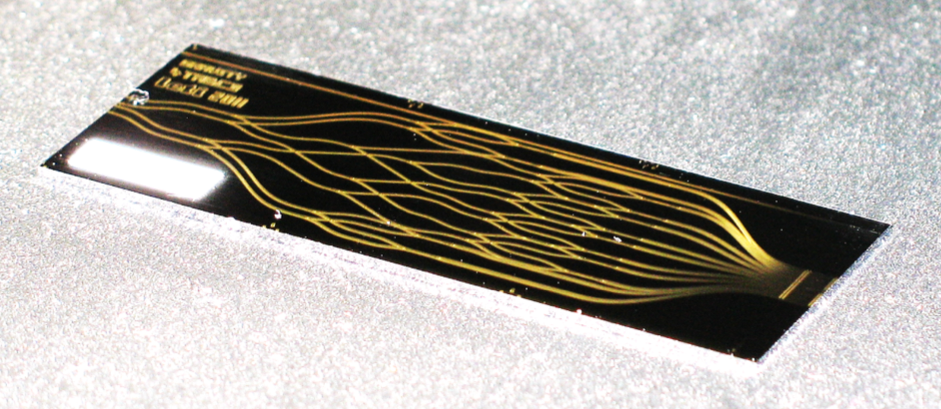}\hspace{0.2cm}
\includegraphics[width=0.4\textwidth]{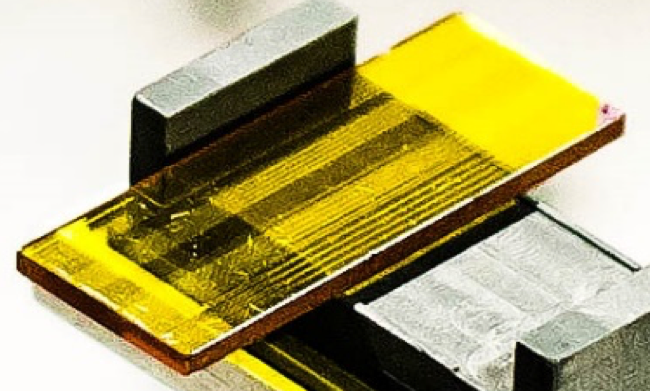}
\caption{Left: view of the current GRAVITY beam combiners and its waveguide network. Courtesy of Laurent Jocou, IPAG. Right: Gallium Lanthanum Sulphide (GLS) chip containing two-telescope directional couplers for the L- and M- bands. Photo J. Tepper.}\label{Fig8}
\end{figure}

A powerful application of photonics in astronomy has been initiated with the technique of pupil remapping, which is an evolution of the technique of aperture masking pioneered by Buscher 1990 and expanded by Tuthill (2000)\cite{Tuthill2000}\,. 
In aperture masking, the non-redundancy of the interferometric sub-array is essential to recover an unperturbed measurements of the closure phases for image reconstruction. This technique has produced astrophysical images with exquisite fidelity and $\lambda$/(2B) resolution, where B is the maximum baseline of the sub-array. The power of this technique with respect to classical adaptive optics is the superior high contrast achievable at a fraction of $\lambda$/D. 
A well-known limitation of aperture masking is nonetheless that the condition of non-redundancy of the array results in using only a small fraction of the pupil, typically between 1\% and 10\% of the incoming photons, biasing the technique to the brightest sources.\\
To circumvent this limitation, Perrin (2006)\cite{Perrin2006} proposed to exploit the so-called pupil remapping technique, a derivation from the aperture masking technique. It is based on using an optical photonics device such a single-mode fiber bundle to transform Ð or remap Ð the fully redundant input pupil of the telescope into a non-redundant output pupil, and extract the non-redundant interferometric visibilities and closure phases from which to reconstruct a high-fidelity image. In practice, the difficulty to calculate a fully non-redundant pupil for more than 15 sub-apertures and the large number of encoding pixels needed (with consequent high readout noise) has limited as well this approach to relatively bright sources. An example of such a fibered single-mode pupil-remapping instrument is presented in Huby (2012, 2013)\cite{Huby2012,Huby2013} with the FIRST instrument operating in the visible range. The instrument was installed after an adaptive optics layer to improve the coupling of the sub-apertures into the single-mode fibers in order to enhance the sensitivity. However, the poor stability of optical fibers with respect to temperature and gravity variations has made the calibration and usage of the instrument non trivial. In order to mitigate the influence of the environmental factors on the photonic remapper, 
Jovanovic (2012) has exploited the advantage of a one-piece remapping device obtained by laser inscription into
a boro-aluminosilicate glass substrate and constituting the core of the Dragonfly instrument. This prototype was successfully tested on-sky to recombine in a multi-axial fashion four sub-apertures of the Anglo-Australian Telescope. \\
In the near-future the performance of pupil remapping instruments will probably be increased by interfacing the remapping function with integrated optics beam combiners similar to the one installed in Gravity (see Fig.~\ref{Fig8}).

\subsection{Nulling interferometry}

Nulling is another interferometric application where photonic devices and fibers can play an important role. 
Mennesson et al. (2002)\cite{Mennesson2002} have shown that using single-mode mid-infrared fibers in a nulling interferometer enables efficient modal filtering of the wavefront and relaxes the constrain on the wavefront quality for the incoming beams by a factor of $\sim$100. This solution was unfortunately not implemented on the Keck nulling interferometer because of the lack of suitable mid-infrared single-mode fibers at the time of construction\cite{Serabyn2012}\,. 
The need for mechanically and thermally stable miniaturized and achromatic beam combiners raised interest in the 2000--2010 decade for the development of mid-IR integrated optics solutions capable of combining two to four telescopes in nulling mode.\\
A very elegant solution was proposed by Wehmeier et al. (2004)\cite{Wehmeier2004} with the concept of ``Magic Tee coupler" imported from the field of microwave engineering down to the mid-IR range. Due to its single-mode design, this device would have provided a fully achromatic destructive output port and a complementary constructive port. However, while excellent single-mode and single-polarization modal filters could be manufactured\cite{Labadie2007}\,, the practical realisation of a full combiner was hampered by the inherent poor throughput of such mid-IR waveguides over propagating distances larger than 10 to 100\,$\lambda$\cite{Labadie2006}\,.\\
At shorter wavelengths, Weber (2004)\cite{Weber2004} demonstrated the potential of integrated optics for nulling achieving stable $\sim$10$^{-4}$ null at 1.5\,$\mu$m over 5\% bandwidth using reverse Y-junction. These results were not further improved due to the fading out of the nulling activities in Europe. At a longer wavelength of 3\,$\mu$m, Heidmann (2012)\cite{Heidmann2012} and Mart\`in (2014)\cite{Martin2014} investigated the potential of active LiNbO$_3$ integrated optics beam combiners to lock the fringes for high-contrast mid-infrared interferometry. More recently, the nulling approach has been revived for instance in the context of closure phase nulling\cite{Chelli2009} and for which possible photonic-based approaches are under investigation\cite{Norris2016}\,.

\subsection{Astrophotonics applications for high-resolution spectroscopy}
In parallel to the effort conducted in the field of interferometric imaging, research groups mainly in the United Kingdom and Australia have extensively focused on taking advantage of the beam-shaping and reformatting properties of photonics to explore new directions for designing high-resolution spectrographs.  
Multimode fibers permits thanks to their large core an efficient coupling of a seeing-limited telescope point spread function. However, they also exhibit modal noise, which introduces spectral line shifts and decrease the overall spectral resolution\cite{Haynes2014}\,. This effect becomes even more dominant at longer wavelengths, for instance in the H-band with respect to the visible range, as the number of modes decreases and the overall output phase scrambling is reduced. Different approaches are considered in this context.

\subsubsection{Photonic Lanterns\cite{Birks2015}}

It is well known that for a slit spectrograph operating in seeing-limited mode, the size of the spectrograph -- or the diameter of the collimator -- must increase proportionally to the diameter of the telescope if more flux is to be collected at an unchanged spectral resolution. This scaling relation breaks if the entrance point spread function of the spectrograph is diffraction limited. The idea of the photonic lantern\cite{LeonSaval2005,Thomson2011} is precisely to adiabatically transform a multimode PSF into many single-mode PSF that can be rearranged in the form of a pseudo-slit prior to spectral dispersion. The advantage is to benefit from a diffraction-limited rearranged slit in the dispersion direction, which removes the limitation of the scaling relation and suppresses modal noise by virtue of the single-mode nature of the pseudo-slit\cite{Harris2015}\,. This principle has been verified on-sky, but requires further assessment on a real high-resolution spectrograph. Also, multimode/single-mode/multimode transitions allow embedded photonic function and retrofit to existing multimode systems.

\subsubsection{Integrated Photonic Spectrographs (IPS) and Fourier Transform Spectrographs (SWIFTS)}

The potential of photonic lanterns has been further observed in the development of fully integrated spectrographs based on the implementation of arrayed waveguide gratings (AWG)\cite{Cvetojevic2012}\,. A seeing-limited PSF is sampled through an adiabatic photonic lantern, which single-mode outputs feed an AWG device. The proof-of-concept was demonstrated on-sky with a preliminary resolving power $R$=2500. \\
Other groups have explored the option of integrated Fourier Transform Spectrographs based on a stationary and counterpropagative system of fringes inside singlemode waveguide. The fringe pattern is sampled and read thanks to regularly spaced nanodetectors in contact with the evanescent part of the field at the surface of the waveguide\cite{Lecoarer2007}\,. A resolving power of R$\sim$350 was obtained due to an under-sampled fringe packet, without further on-sky demonstration. Following this first success, this concept is being now exploited for commercial applications in the visible and near-IR range where a resolving power up to $R$=150,000 is reached\footnote{See http://resolutionspectra.com/products/zoom-spectra}. 

\subsubsection{Suppression of sky emission lines}
A very interesting application of multimode to single-mode photonic lantern is seen in the field of near-IR spectroscopy. Indeed, the delivery of a set of single-mode output by a lantern can also be used to feed a batch of single-mode Fiber Bragg Grating (FBG) wavelength filters. 
Bragg gratings are optical devices which present periodic variations of their refractive index and which fulfill the high-reflectivity Bragg condition for specific wavelengths over a narrow bandwidth. When such periodic index variations are applied along an optical fiber, it is possible to transmit a spectrum alleviated from unwanted wavelengths, which are reflected back. This principle was implemented into aperiodic Fiber Bragg Gratings designed to suppress 94\% of the OH sky emission lines in the range 1.45 to 1.6\,$\mu$m\cite{BlandHawthorn2011}\,. Here again, photonics offers a unique solution to mitigate a well-known issues in observational astrophysics. 

\subsection{Active Integrated Optics}

Electro-optic materials where the refractive index can be modified by applying an external electric field and where optical waveguides can be realized have since a long time been used in the telecommunications for optical routing, intensity modulation and optical phase delay applications. Using these concepts for astrophotonics, opens the way for on-chip photometry balancing, fringe scanning and fringe locking, in a wide range of wavelengths. One of the most popular materials for active integrated optics is Lithium Niobate, where high contrast (36dB) rejection ratios have been obtained in the mid-IR\cite{Martin2014}\,. The transparency range of this material allows to cover from mid-infrared (L-band)\cite{Hsiao2009} down to visible\cite{Martin2016}\,.
 One of the main issues concerning interferometry applications in astronomy is the chromatic dispersion of the fringes and the high propagation losses in the mid-IR due to low field confinement. Electro-optics can here be used to compensate for dispersion by cascading the electrodes and slightly modifying the refractive index of the waveguides. Finally, recent results on direct laser writing in similar electro-optic materials allows to access the 3D fabrication of waveguides, thus increasing compacity and avoiding crossings\cite{He2013}\,. 


\section{Importance of fibers}

\noindent Nowadays, optical fibers play a central role in the development phase of any astrophotonic instrument. In particular they allow the efficient interconnection between the telescope beam(s) and the type of integrated optics devices or functions in which this beam is to be coupled in. They permit to bypass classical fore-optics/beam-resizing units and they have demonstrated to significantly simplify the design of multi-aperture interferometers. They also permit in principle to transport beams over a long distance from the telescope focus down to an optical back-end. 
Experiments such as the OHANA project have attempted to bypass complex optical trains of the Keck interferometer using long ($\sim$300 m) single-mode fiber cables directly plugged at each telescopeÕs focus\cite{Perrin2006b}\,. 
Multimode fibers in fiber-fed spectrographs efficiently enable multiplexed observations of few hundreds objects per pointing. Current examples are the visible high-resolution spectrograph PEPSI on the LBT\cite{Strassmeier2015} or the near-IR fiber multi-object spectrograph FMOS on Subaru\cite{Kimura2010}\,. \vspace{0.25cm}
\\
The technicals requirements are in a large part similar to those presented in Sect.~\ref{Sect22}: highly transparent material in the spectral range of interest; propagation losses as low as few dB/km for {\it beam transportation} and few dB/m for {\it beam combination} (i.e. when the propagation lengths involved are in the order of 0.5\,cm); single-mode behaviour for wavefront filtering or multimode behaviour to improve throughput\cite{Tatulli2010}\,; numerical aperture compatible with a large variety of fast and slow injection optics to minimize insertion losses; controlled high-birefringence for polarization-maintaining fibers in interferometric applications, or at the contrary polarization scrambling for applications where modal noise needs to be mitigated. The difference to integrated optics is the requirement on the technological platform since the fabrication process is quite different: fibers are manufactured through extrusion, cooling and stretching process that should however limit as much as possible local inhomogeneities in the propagation medium to avoid unwanted scattering losses. We summarize hereafter some properties of fibers that can be used in astrophotonics instruments.

\subsection{Visible and near-infrared silica fibers}
Optical fibers based on silica glasses (SiO$_{\rm 2}$) have been extensively developed for the field of telecommunications over the last forty years. The enormous effort -- and subsequent market -- behind this technology has permitted a constant improvement in the fibers performance in terms of transparency, polarization control or mechanical reliability. At the telecom wavelength of 1.55$\mu$m, the propagation losses of modern fibers are of the order of $\sim$0.2\,dB/km (i.e. an attenuation by a factor $\sim$0.05 per kilometer)\footnote{The first original fiber by Kao \& Hockham had propagation losses of 1000\,dB/km.}. The absorption of the silica glass itself can be as low as 0.02\,dB/km, but the presence of sub-micron scattering centers resulting from the fabrication process reduces the homogeneity of the medium and induces scattering losses. In silica fibers, Rayleigh scattering becomes a dominant effect towards shorter wavelengths than 1.5$\mu$m. At wavelengths longer than 1.5$\mu$m infrared molecular absorption bands (mainly due to OH$^{-}$ hydroxyls) become stronger. The combination of these two effects is minimized around 1.5-1.6$\mu$m. While being too lossy for the telecom market beyond 1.9$\mu$m, silica fibers still exhibit some decent ($\sim$50-60\%) transmission for astronomical applications in the K band \cite{Laurent2002} for few meters length, but become significantly opaque beyond $\sim$2.2$\mu$m.\\

\subsection{Growth of the mid-IR fibers market}

The first mid-infrared fibers were fabricated in the mid-sixties from arsenic-sulphur glasses \cite{Kapany1965} and exhibited losses of circa 10\,dB/m. Since then numerous groups have attempted to develop high-throughput and reliable mid-infrared fibers. The availability of highly transparent infrared glasses and crystals is large, but the fabrication process is delicate and mid-infrared fibers acquired a reputation of being fragile, difficult to handle, and in some cases toxic when arsenic (As) is employed. 
The market volume for mid-infrared fibers has grown slower in comparison to what was observed in the telecom field, and no dominant and fast-growing technological platform has emerged. Primarily driven by chemical sensing, medical or military applications, the scarcity of interoperable hardware (e.g. laser diodes, IR couplers, cheap uncooled detectors...) and the high cost of infrared optics hindered the large-scale development of mid-infrared fibers. \\
The situation has significantly evolved in the last ten years with a sustained and general growth of the mid-infrared photonics field, which has made available a larger number of components with excellent performances and lower price by most traditional suppliers of laboratory hardware. This is also beneficial for the field of experimental astronomy. 

\subsubsection{Dielectric fibers}

\paragraph{Fluoride-glass fibers} Fluoride glasses are oxide-free glasses based on fluoride compound. They were discovered in the seventies at the University of Rennes \cite{Poulain1977}\,. Fluorozirconate ZBLAN glasses (ZrF$_{\rm 4}$:BaF$_{\rm 2}$:LaF$_{\rm 3}$
:AlF$_{\rm 3}$:NaF) are the best-known of the so-called Heavy Metals Fluoride Glasses (HMFG). The theoretical intrinsic losses of ZBLAN in the infrared could be as low 0.001 -- 0.01\,dB/km, which makes them extremely transparent even in comparison to silica. Fluoride glasses therefore attracted a significant interest from the side of fibers manufacturers. However, the formation of impurities in the material during the manufacturing process results in additional scattering, which typically brings the propagation losses of fluoride fibers to $\sim$1\,dB/km. The transparency range in the infrared extends up to $\sim$5--6$\mu$m, covering the L and M bands. Compared to other non-oxide infrared glasses, fluoride fibers have a relatively low core index ($n\sim$1.5), which minimizes the losses due to Fresnel back reflections. ZrF$_{\rm 4}$ and InF$_{\rm 3}$ fluoride fibers are now commercially available at an accessible price by companies such as Thorlabs in both large-core multimode and small-core single-mode versions for the 3--5$\mu$m range. A drawback of this type of fibers compared to silica fibers is their sensitivity to crystallization effects and moisture, in addition to their mechanical fragility. For instance, the maximum bending radius of such fibers is on the order of $\sim$20\,cm, which makes them less flexible than silica fibers. In metrology application, Raman scattering peaks can result in unwanted spurious flux in the system (F. Eisenhauer, private communication). Due to their superior transparency compared to silica fibers between 2 and 2.4$\mu$m, fluoride fibers have been used in K-band interferometry. 

\paragraph{Chalcogenide-glass fibers} Another class of non-oxide glasses used to manufacture mid-infrared fibers are Chalcogenide glasses, which is a generic description of materials composed of elements such as Arsenic (As), Germanium (Ge), Sulfur (S) or Selenium (Se). The glass infrared transparency is more extended than for fluoride glasses: it reaches up to $\sim$12$\mu$m and can rise to $\sim$18$\mu$m by doping with heavy compounds such as Tellurium (Te) or Silver.
Some classical compositions are Arsenic trisulfide (As$_{\rm 2}$S$_{\rm 3}$) or triselenide (As$_{\rm 2}$Se$_{\rm 3}$), which are however potentially toxic due to the presence of Arsenic. 
The average refractive index is circa $\sim$2.3--2.5 in the mid-infrared, i.e. significantly higher than for fluoride and silica glasses. These are chemically stable glasses and unaffected by moisture. In the 3--5$\mu$m range strong absorption bands due to H$_{\rm 2}$O and OH$^{-}$ contaminants can degrade the transparency of the material, making a strong purification necessary during the glass synthesis. 
Chalcogenide As-S fibers have been manufactured in the 2000s at the Naval Research Laboratories with propagation losses measured to 0.1--1 dB/m in the 2--6$\mu$m spectral range\cite{Aggarwal2002}\,. Samples of 20-cm long {\it single mode} fibers were produced to serve as modal filters in astronomical interferometry applications \cite{Ksendzov2007}\,. With a 23-$\mu$m core they were tested at 10.6$\mu$m and presented propagation losses of $\sim$8\,dB/m and a total transmittance of 43\% accounting for the Fresnel losses. 
Both multimode and single mode chalcogenide fibers are now commercially available by main suppliers like Newport or CorActive with core diameters ranging from 6$\mu$m (with a 2--4$\mu$m transparency range) to 800$\mu$m (and a transmittance up to 9$\mu$m. For the large-core fibers, the average propagation losses can be as low as 0.2--0.3\,dB/m up in the 2--7$\mu$m range. As most of the IR fibers, chalcogenide fibers are more fragile than silica-based fibers and can be easily broken. To the best of our knowledge, no chalcogenide single-mode fiber was ever used on-sky to perform modal filtering or interferometric beam combination in the mid-infrared.

\subsection{Hollow air-core fibers}\label{hollowcore}

\noindent An alternative to dielectric materials to guide infrared radiation are the hollow air-core fibers. They are core-less -- or air-core -- waveguides surrounded by a higher index dielectric film plus a high-reflectivity metallic layer (e.g. Aluminium), which then behave as "leaky" waveguides (cf. Fig.~\ref{hollowcore}). Because the core medium is air, the propagation of the guided mode is theoretically lossless and unaffected by chromatic dispersion. However the transmission of hollow core fibers is practically hampered by the level of surface roughness of the inner walls coating, which implies scattering losses. Hollow core waveguides were investigated for several years by J.~A. Harrington and his team at Rutgers University and nowadays flexible multimode hollow core waveguides with core diameters of 500\,$\mu$m or larger are commercially available for CO2 or Er:Yag laser power delivery\footnote{The manufacturer {\it Molex -- Polymicro Technologies} has been for years a leader in the fabrication of hollow core fibers.}. Typical losses for multimode waveguides are measured between 0.5 and 1\,dB/m. Additional losses have to be considered depending on the cross-sectional size of the fiber. Indeed, the waveguide attenuation factor $\alpha$ increases proportionally to 1/$a^3$, where a is the radius of the fiber core. Hence the smaller the core, the larger the attenuation. This has a strong incidence on the viability of {\it single mode} fibers with small cores of the order of the operating wavelength, and for which propagation losses would amount to tens of dB/cm.

\subsection{Microstructured fibers and multicore fibers}\label{microstructured}
\begin{figure}[t]
\centering
\includegraphics[width=6.2cm]{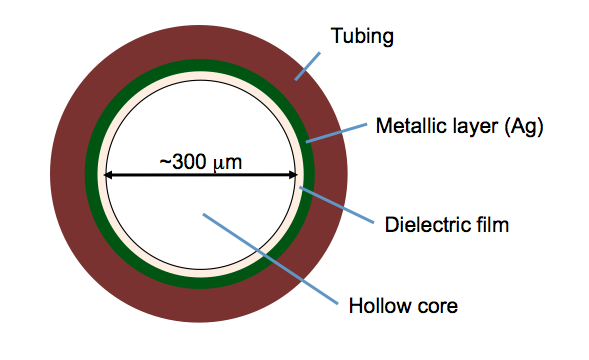}
\includegraphics[width=5.2cm]{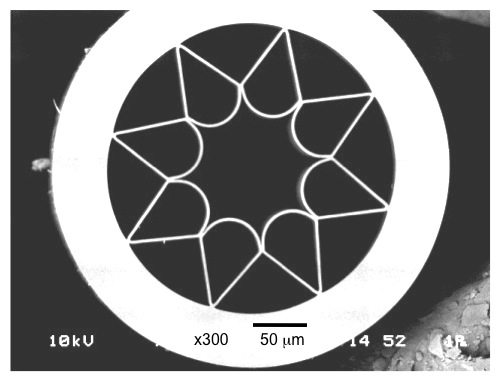}
\includegraphics[width=5.1cm]{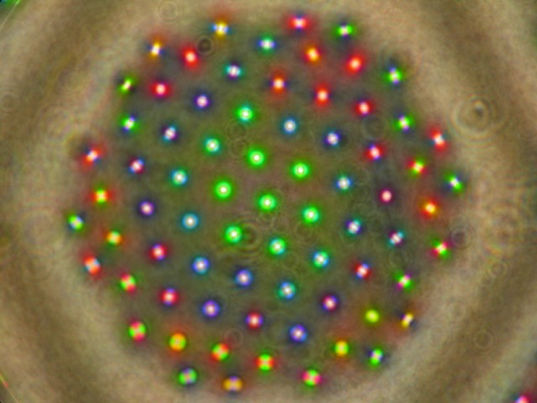}
\caption{{\it Left:} sketch of a multimode hollow-core fiber. Adapted from Labadie (2009)\cite{Labadie2009}\,. {\it Middle:} Cross section of Kagome PCF for the 3$\mu$m range\cite{Yu2012}\,. 
{\it Right:} A multicore section of a early prototype scrambling multicore fibre design with 73 cores. The different colours and mode patterns are indicative of different core diameter, designed to scramble the phase of light waves and so reduce the fibre modal noise.}\label{Fig9}
\end{figure}

\noindent So far the type of fibers presented are based on the core/cladding geometry. Microstructured fibers are a new and alternative approach to this basic geometry, with the best known example being the photonic crystal fibers (PCFs) developed now for circa 20 years. PCFs were invented in the mid nineties at the University of Bath in the group of Russell, Knight and colleagues \cite{Knight1996,Russell2003}\,. The major advantage of PCFs is their flexibility in terms of core design, which in return permits to taylor the macroscopic properties of the fiber (field confinement and modal behaviour, mode diameter) with great latitude depending on the targeted application. Photonic crystal fibers consist in dielectric structures with periodically varying refractive index at the wavelength scale thanks to regularly spaced air holes. The two classes of photonic crystal fibers are 1) the index-guiding PCFs, where a high-index core is surrounded by a region of same material and an effective index lowered by the presence of the holes. 2) the photonic band gap (PBG) guiding PCFs that present an additional low-index hole or defect at the centre of the holes arrangement, and for which light guidance is ensured by the bandgap effect used in Bragg mirrors and fibers. Typically the photonic band gap region is spectrally very narrow.\\
A powerful aspect of index-guiding PCFs is their potential "endlessly" single mode behaviour, which can be formally obtained at any wavelength with photonic crystal fibers. The normalized $V$ frequency, which must verify $V$$<$2.405 for single mode behaviour in classical step index fibers, is defined for index-guiding PCFs by $V$=$(2\pi\Lambda/\lambda)\sqrt{n_{\rm 0}^{2}-n_{\rm {eff}}^{2}}$, where $\Lambda$ is the center-to-center holes spacing. In contrast to classical step-index fibers, $V$ approaches a constant value for large $\Lambda$/$\lambda$. This means that by proper choice of the hole diameter to hole spacing ratio $d$/$\Lambda$ a single mode behaviour is observed at any wavelength \cite{Birks1997}\,. A second important aspect of microstructured fibers is their {\it large mode area} (LMA) properties even for single mode waveguides, which allows to couple telescope beams with different numerical apertures. \\
\noindent Most PC fibers currently developed are based on silica technology and therefore well adapted to the visible and near-infrared range. The overall propagation losses have been measured to tens of dB/km at 1550\,nm, but with continuing improvement and new low-loss records of few dB/km \cite{Russell2006}\,. Since few years, commercial PCFs can be acquired by major optical suppliers like Thorlabs \footnote{Manufactured by the company NKT Photonics.}, who offer endlessly single mode LMA fibers, Photonic Bandgap fibers, or polarisation maintaining fibers transmitting up to the 2.1$\mu$m, near the silica cutoff. 
In the mid-IR range, pioneering works have been achieved by various research groups to extend the potentialities of microstructured fibers to wavelengths up to 10$\mu$m\cite{Temelkuran2002}\,. 
\begin{figure}[t]
\centering
\includegraphics[width=7.0cm]{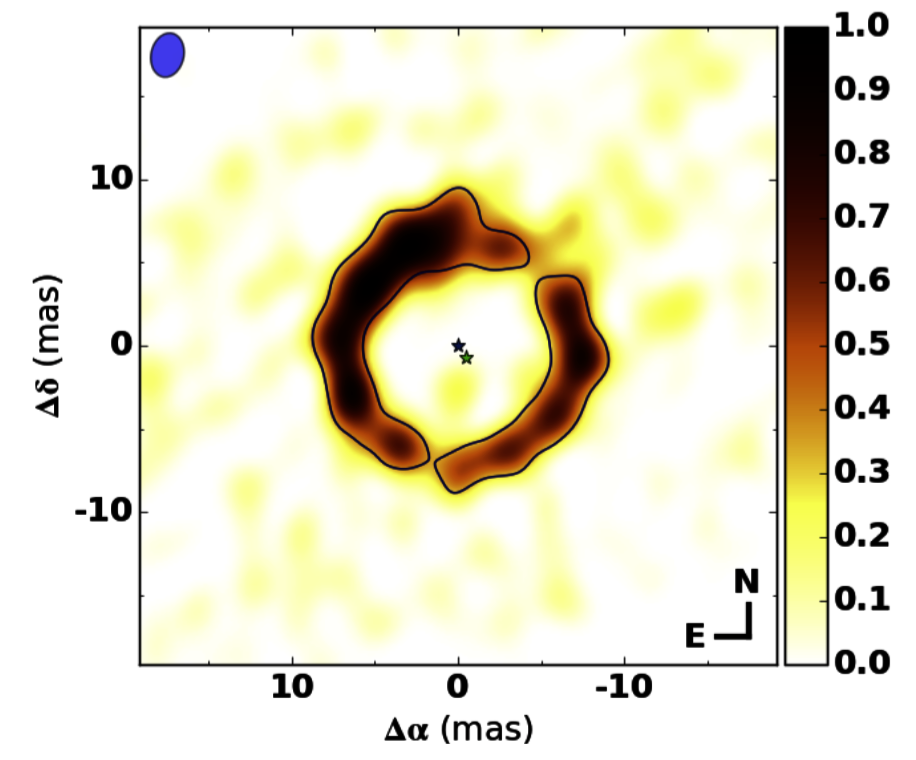}\hspace{0.25cm}
\includegraphics[width=7.0cm]{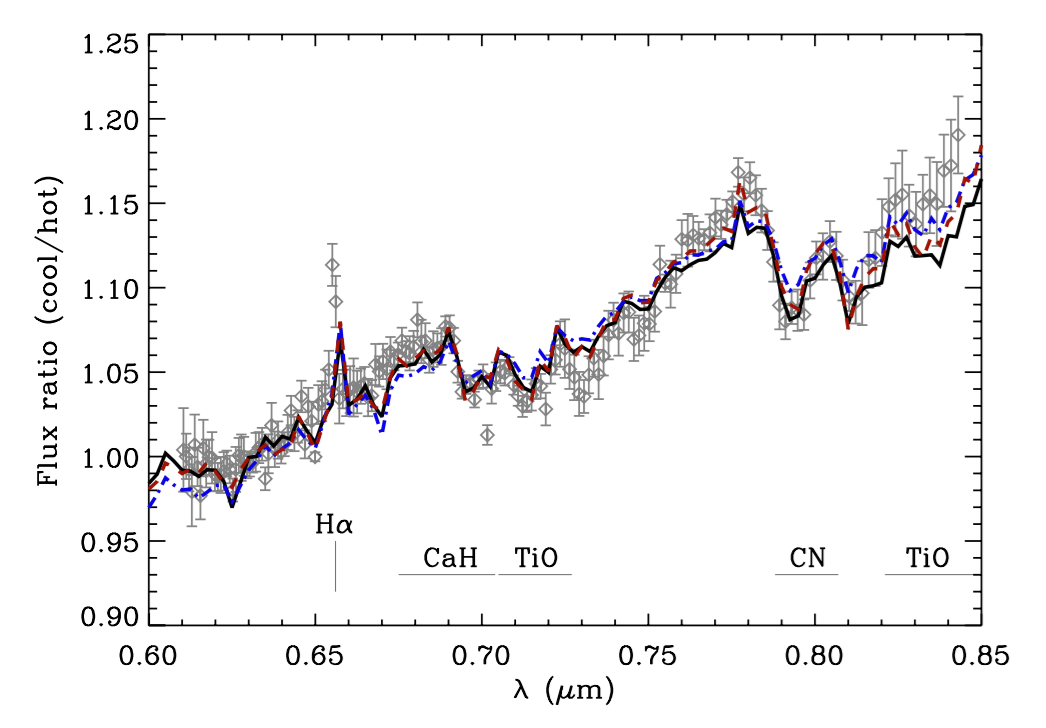}
\caption{{\it Left:} interferometric imaging in the H-band with PIONIER of the dust sublimation front of a post-AGB circumbinary disk\cite{Hillen2016}\,. {\it Right:} Fractional spectrum of the Capella system obtain with FIRST in the visible with the main atomic and molecular features\cite{Huby2013}\,.}\label{Fig10}
\end{figure}
A promising approach has been developed for the 3--4$\mu$m domain with new silica-based microstructures fibers. Despite the well-known low transmittance of silica beyond 3$\mu$m, the high stability and fabrication maturity of this material compared to the more traditional chalcogenide glasses and silver halide crystals for mid-IR fibers justify to attempt using the silica platform for light guidance in the L astronomical band. In a recent work, Yu (2012)\cite{Yu2012} explored the properties of microstructured hollow-core(s) photonic crystal fibers based on a Kagome lattice in the 3 to 4$\mu$m spectral range (Fig.~\ref{hollowcore}). Because they are not based on the existence of a photonic bandgap, they also posses a broader spectral bandwidth. The fundamental mode is efficiently confined in the 90$\mu$m core of the fiber, with a similar MFD. Most interestingly, the region between 3 and 3.25$\mu$m shows a clean and almost flat minimum of the attenuation curve down to 0.05\,dB/m,  which is a factor 5 to 10 better than the small-core single mode fluoride fibers commercially available. Such solutions can result to be very promising for single-mode astronomical applications in the 3--5$\mu$m range.\\
In the field of high-resolution spectroscopy, multicore fibers appear as a very promising route to combine the coupling efficiency of a multimode fiber with the modal stability of a single-mode structure. Several core geometries have been investigated\cite{Haynes2014} and an example is shown in Fig.~\ref{Fig9}. Multicore fibres (MCF) essentially allow multiple single mode, few mode or highly multimode operation within a single fibre structure. MCF devices can be tapered down at the input and/or output to form photonic lanterns structures\cite{Birks2015} that if suitable mode matched can provide efficient conversion from a highly multimode regime to the single or few mode regime. In the field of high-resolution spectroscopy, multicore fibres appear as a very promising route to combine the coupling efficiency of a multimode fibre with the modal stability of a single-mode structure. Several core geometries have been investigated and an example is shown in Fig. 9. In addition MCF where initially developed as low cost option for suppression of the sky emission lines\cite{Lindley2014} using fibre Bragg gratings in which a FBGs are written in a single operation and packaged as a single compact unit.

%
%
%
\section{Perspectives and conclusions}

We reviewed in this paper part of the fundamental principles of astrophotonics, the pertinent scientific and technical requirements, and present a state of the art of the field, mainly focusing on the cases of astronomical interferometry and high-resolution spectroscopy. Great successes have been obtained in the last decade delivering unprecedented astrophysical results, in particular in fields were high-angular resolution is the main driver. Examples can be cited for instance with the imaging capabilities of the four-telescope PIONIER instrument at the VLTI or the FIRST pupil remapping imager (see Fig.~\ref{Fig10}). Several concepts and technologies are still in the growing phase and will possibly become operational for astronomy in the next decade. The potential for space astronomy is certainly large given the reduced sizes and masses, however a long process of technological consolidation is still forward to meet the TRL criteria driving space technologies.\\
Astrophotonics appears to be a vivid and rapidly growing field: the synergies with astrophysics remain still largely unexploited and new astrophysics-driven paths are expected to be open and expanded in the future.



 

\bibliography{report}   
\bibliographystyle{spiebib}   

\end{document}